\documentclass[12pt]{aastex6}

\usepackage{amsmath}
\usepackage{multirow}
\usepackage{url}
\usepackage{mathptmx}
\usepackage{graphicx}

\shorttitle{Evidence of a gamma-ray spectral break of the LMC}
\begin{document}
\title{Evidence of a spectral break in the gamma-ray emission of the disk component of Large
Magellanic Cloud: a hadronic origin?}
\author{Qing-Wen Tang$^{1,3}$, Fang-Kun Peng$^{2,3}$, Ruo-Yu Liu$^{4}$, Pak-Hin Thomas Tam$^{5}$ and Xiang-Yu Wang$^{2,3}$ }
\affiliation{$^{1}$School of Science, Nanchang University,
Nanchang 330031, China}
\affiliation{$^{2}$School of Astronomy and
Space Science, Nanjing University, Nanjing 210093, China}
\affiliation{$^{3}$Key laboratory of Modern Astronomy and
Astrophysics (Nanjing University), Ministry of Education, Nanjing
210093, China}
\affiliation{$^{4}$Max-Planck-Institut f\"ur Kernphysik, 69117
Heidelberg, Germany}
\affiliation{$^{5}$School of Physics and Astronomy,
Sun Yat-Sen University, Zhuhai 519082, China}

\begin{abstract}
It has been suggested that high-energy gamma-ray emission
($>100{\rm \ MeV}$) of nearby star-forming galaxies
may be produced predominantly by cosmic rays colliding with the
interstellar medium through neutral pion decay. The pion-decay
mechanism predicts a unique spectral signature in the gamma-ray spectrum, characterized by a fast rising spectrum (in $E^2F(E)$ representation) and a spectral break below a few hundreds of MeV. We here report the evidence of a spectral break around 500~MeV in the disk emission of Large
Magellanic Cloud (LMC), which is found in the analysis of the gamma-ray data extending down to 60 MeV observed by {\it Fermi}-Large Area Telescope. The break is well consistent with the pion-decay model for the gamma-ray emission, although leptonic models, such as the electron bremsstrahlung emission, cannot be ruled out completely.
\end{abstract}


\section{Introduction}

It is generally believed that Galactic cosmic rays
are accelerated by supernova remnant (SNR) shocks~\citep{1964ocr..book.....G}. Cosmic-ray (CR)
protons  interact  with  the interstellar gas and produce neutral
pions (schematically written as $p+p\rightarrow \pi^0+{\rm other \, products}$), which
in turn decay into gamma rays. Cosmic-ray electrons can also
produce gamma rays via bremsstrahlung and inverse Compton (IC)
scattering emission~\citep{2010ApJ...722L..58S, 2013ApJ...773..104C, 2015ApJ...808...44F}.
Detailed calculation of the CR propagation in
our Galaxy using the GALPROP code finds that $\pi^0$-decay
gamma rays form the dominant component of the  diffuse  Galactic
emission (DGE) above 100~MeV, while the bremsstrahlung and IC
emissions contribute a subdominant, but  non-negligible
fraction~\citep{2010ApJ...722L..58S}. GeV gamma-ray emissions have  also been detected
from nearby star-forming galaxies~\citep{2010ApJ...709L.152A,2012ApJ...755..164A,2014ApJ...794...26T,2016ApJ...821L..20P,2016ApJ...823L..17G}, and they are interpreted as arising dominantly from cosmic-ray protons
colliding with the interstellar gas as well~\citep{2002ApJ...575L...5P,2004ApJ...617..966T,2007ApJ...654..219T,
2007APh....26..398S,2010MNRAS.403.1569P,2011ApJ...734..107L}.

Although these theoretic arguments favor the pion-decay model for
the GeV gamma-ray emission in these galaxies,  there is no direct
evidence for such pion-decay mechanism. Recently, {\it Fermi}-
Large Area Telescope (hereafter LAT) detected a characteristic
pion-decay feature in the gamma-ray spectrum of two supernova
remnants, IC 443 and W44~\citep{2013Sci...339..807A}.  The pion-decay
spectrum in the usual $E^2F(E)$ representation rises steeply
below several hundred of MeV and then breaks to a softer spectrum. This characteristic spectral feature (often
referred to as the ``pion-decay bump'') uniquely identifies
pion-decay gamma rays and thereby high-energy CR protons.

Motivated by this, we attempt to study the gamma-ray spectra of
nearby star-forming galaxies and examine such  unique pion-decay
bump spectral signature. The LMC is the brightest external galaxies in
gamma-ray emission, as it is very close to us (only 50 kpc). The LMC
is near enough that individual star-forming regions can be
resolved and thus their contribution can be removed  so that one
can obtain a relatively pure diffuse disk component. The high
Galactic latitude of the LMC also leads to a low level of
contamination due to the Galactic diffuse gamma-ray emission.   We
analyze the {\it Fermi}-LAT data of the LMC and pay special attention to the
gamma-ray spectrum extending to 60~MeV using 8 years of {\it Fermi}-LAT Pass 8 data. We
find that the gamma-ray spectrum shows a rise in $E^2F(E)$
representation at low-energies and breaks to a softer spectrum at about 500 MeV.

Our work is different from earlier works
based on the {\it Fermi}-LAT observations of the LMC~\citep{2010A&A...512A...7A,2015ApJ...808...44F,2016A&A...586A..71A},
which focus on the gamma-ray emission above 200~MeV.

\section{Data analysis and results}

\subsection{Data selection}
The LAT Pass 8 data between 2008 August 4 and 2016 August 4 are taken from the {\it Fermi} Science Support Center (hereafter FSSC)\footnote{\url{https://fermi.gsfc.nasa.gov/ssc/}}. Events with energy between 60~MeV and 100~GeV are selected. These data are analyzed using the {\it Fermi}
Science Tools package (v10r0p5) available from the FSSC.
We select ``FRONT+BACK'' SOURCE class events and use instrument response functions
P8R2$\_$SOURCE$\_$V6.
Events with zenith angles $>$90$^\circ$ are excluded to reduce the contribution of
Earth-limb gamma rays as well as that with the rocking angle of the satellite
was larger than $52$ degrees. Gamma rays in a box Region Of Interest (ROI), 20$^\circ\times$20$^\circ$ centering
at the position of RA.=80.894$^\circ$, Dec=-69.756$^\circ$, are used in the spectrum analysis between 60~MeV and 100~GeV, in which energy dispersion correction is considered.
Binned maximum likelihood analysis are performed in this work\footnote{\url{https://fermi.gsfc.nasa.gov/ssc/data/analysis/scitools/binned_likelihood_tutorial.html}}.

\subsection{Background sources}

With a large angular containment of the low energy gamma-ray photons, all identified sources from the third LAT catalog (3FGL, \citet{2015ApJS..218...23A}) within 20$^\circ$ from the center position are included. Four of them are excluded because they are located in the intensive LMC region~\citep{2016A&A...586A..71A}, which correspond to 3FGL J0454.6-6825, 3FGL J0456.20-6924, 3FGL J0525.2-6614 and 3FGL J0535.3-6559. Another point source, 3FGL J0537.0-7113, is also at the edge of the LMC region and thus is excluded. These sources should be removed from the background sources so that the LMC sources can be distinguished significantly. A total of 72 point sources are included with fixed positions as in the 3FGL.

\begin{figure}
\centering
\includegraphics[scale=0.80]{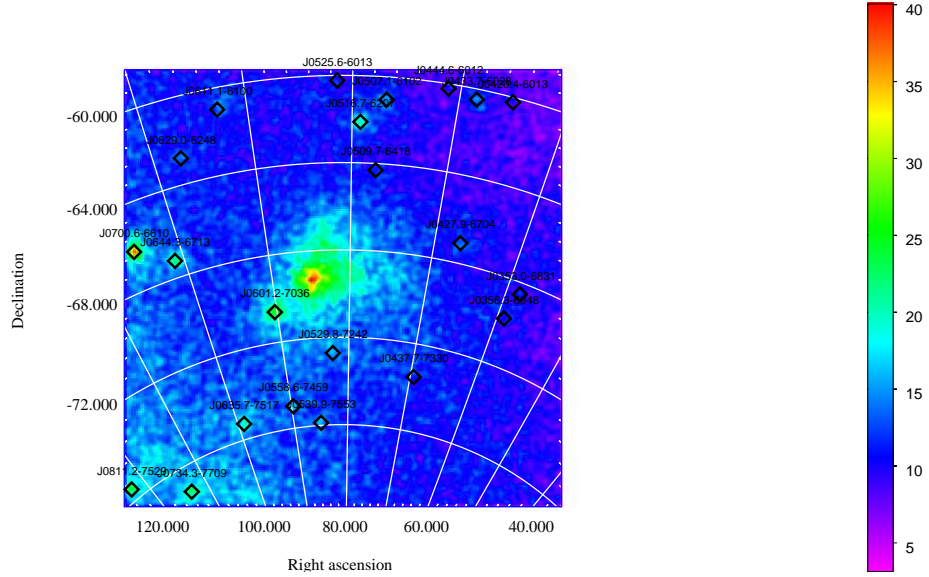}
\caption{ Gamma-ray count map of the
$20^{\circ}\times20^{\circ}$ fields around the LMC in the energy range
from 60~MeV to 2.45~GeV observed by the {\it Fermi}-LAT. The nearby background point sources are marked with black diamonds. Events were spatially binned in pixels of side length $0.2^{\circ}$.\label{fig:f1}}
\end{figure}

For sources within the ROI, the spectral parameters are fixed at the 3FGL catalog values except for their normalisations, which are allowed to be free. A total of 22 point sources are selected that allow the normalisations to be free, which are marked as black diamond in Fig. \ref{fig:f1}. For sources outside of ROI, all spectral parameters are fixed at the 3FGL catalog values.

The Galactic diffuse background and isotropic gamma-ray background are given by the
templates ``gll$\_$iem$\_$v06.fits'' and ``iso$\_$P8R2$\_$SOURCE$\_$V6$\_$v06.txt''
available in the FSSC, while their normalizations are allowed to vary.

\subsection{The LMC sources}
\subsubsection{The LMC point sources and extended sources}
The LMC sources are categorized into two subparts, namely, the point ones and the extended ones.

We include four newly-identified point sources (namely P1, P2, P3, P4) in the LMC field in the analysis, which correspond to PSR~J0540-6919, PSR~J0537-6910/N157B, a gamma-ray binary CXOU~J0536-6735 and N~132D respectively~\citep{2016A&A...586A..71A,2016ApJ...829..105A}. Their positions are determined by~\citet{2016A&A...586A..71A}.

\begin{figure}
\centering
\includegraphics[scale=0.60]{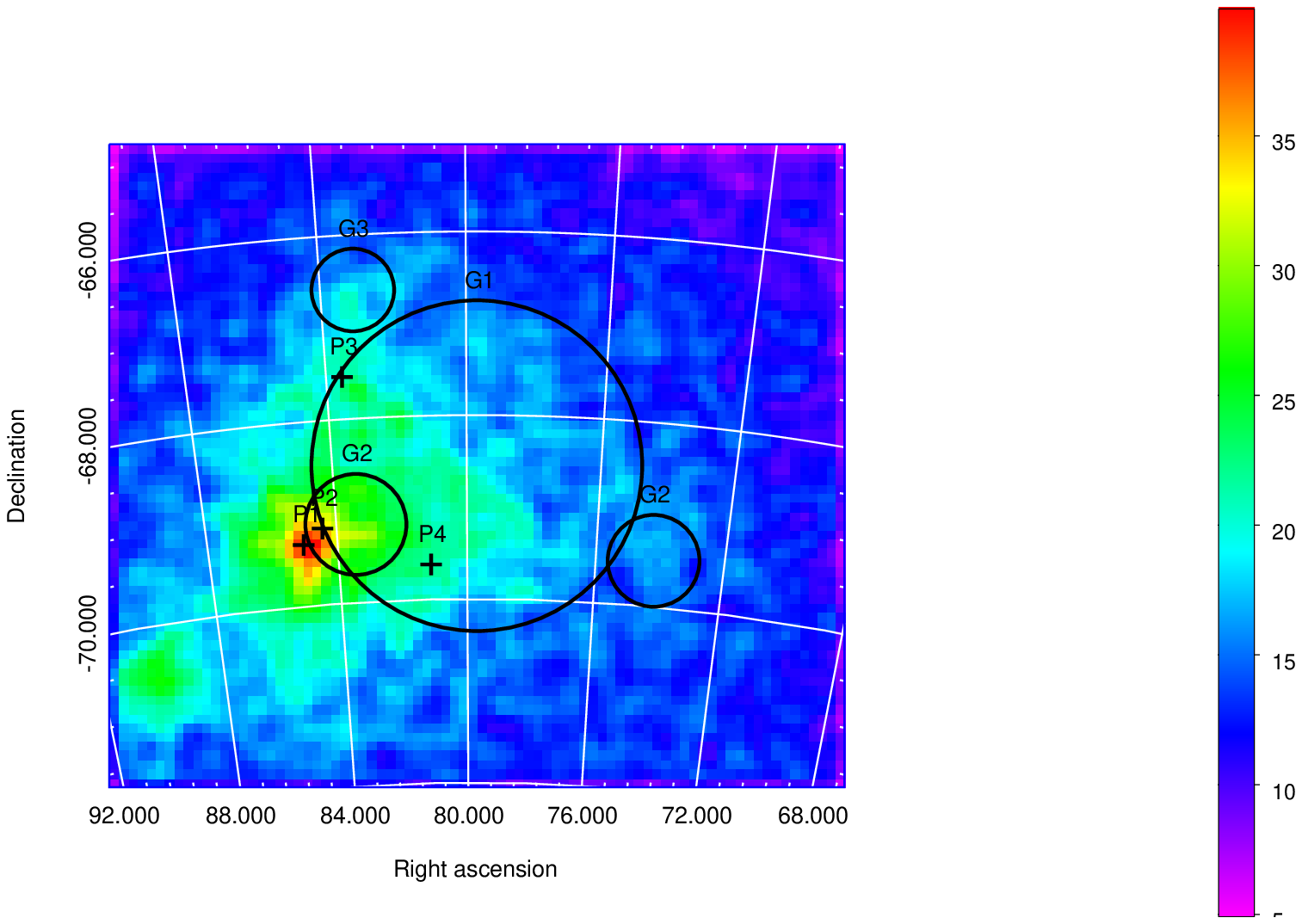}
\includegraphics[scale=0.60]{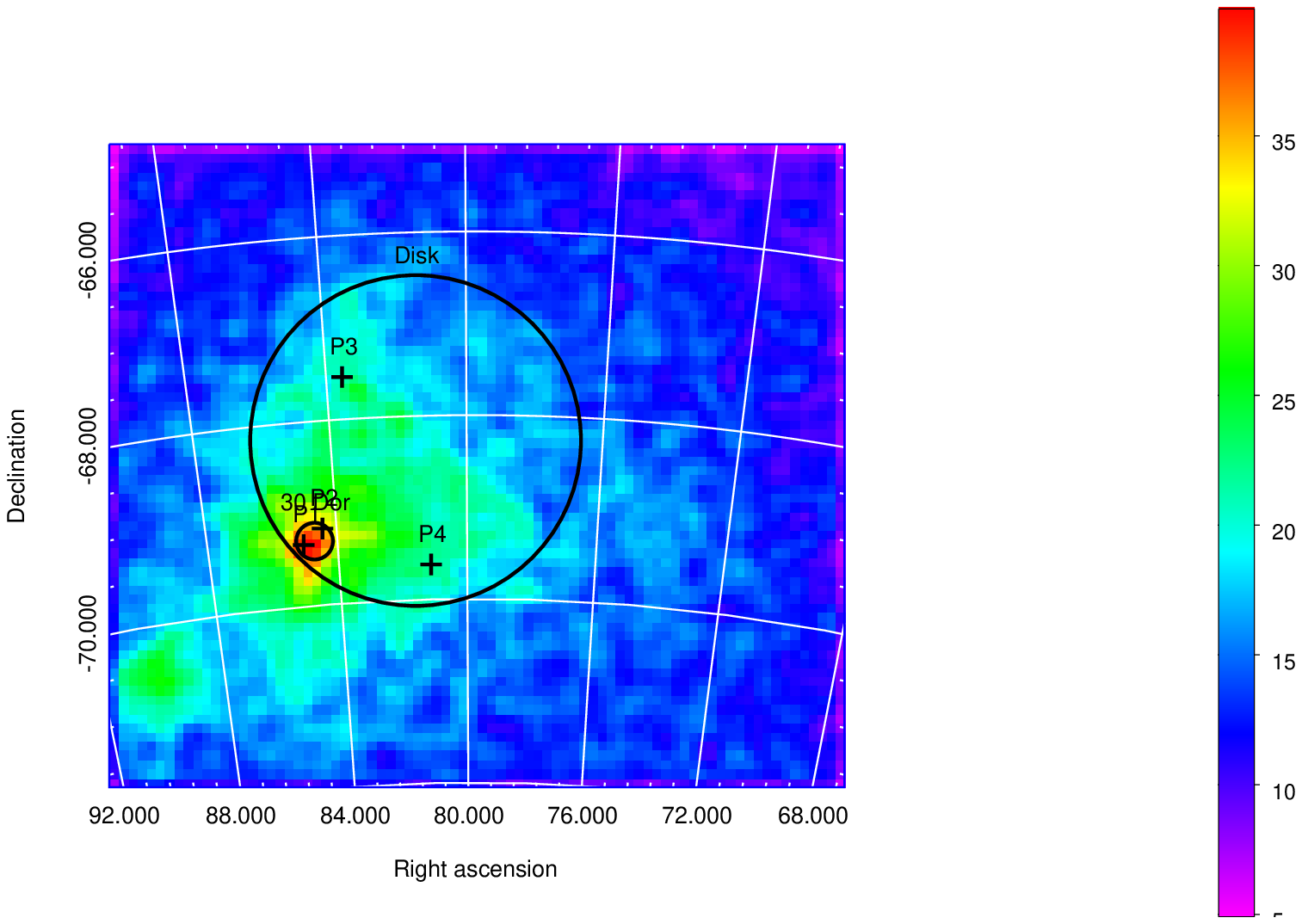}
\includegraphics[scale=0.60]{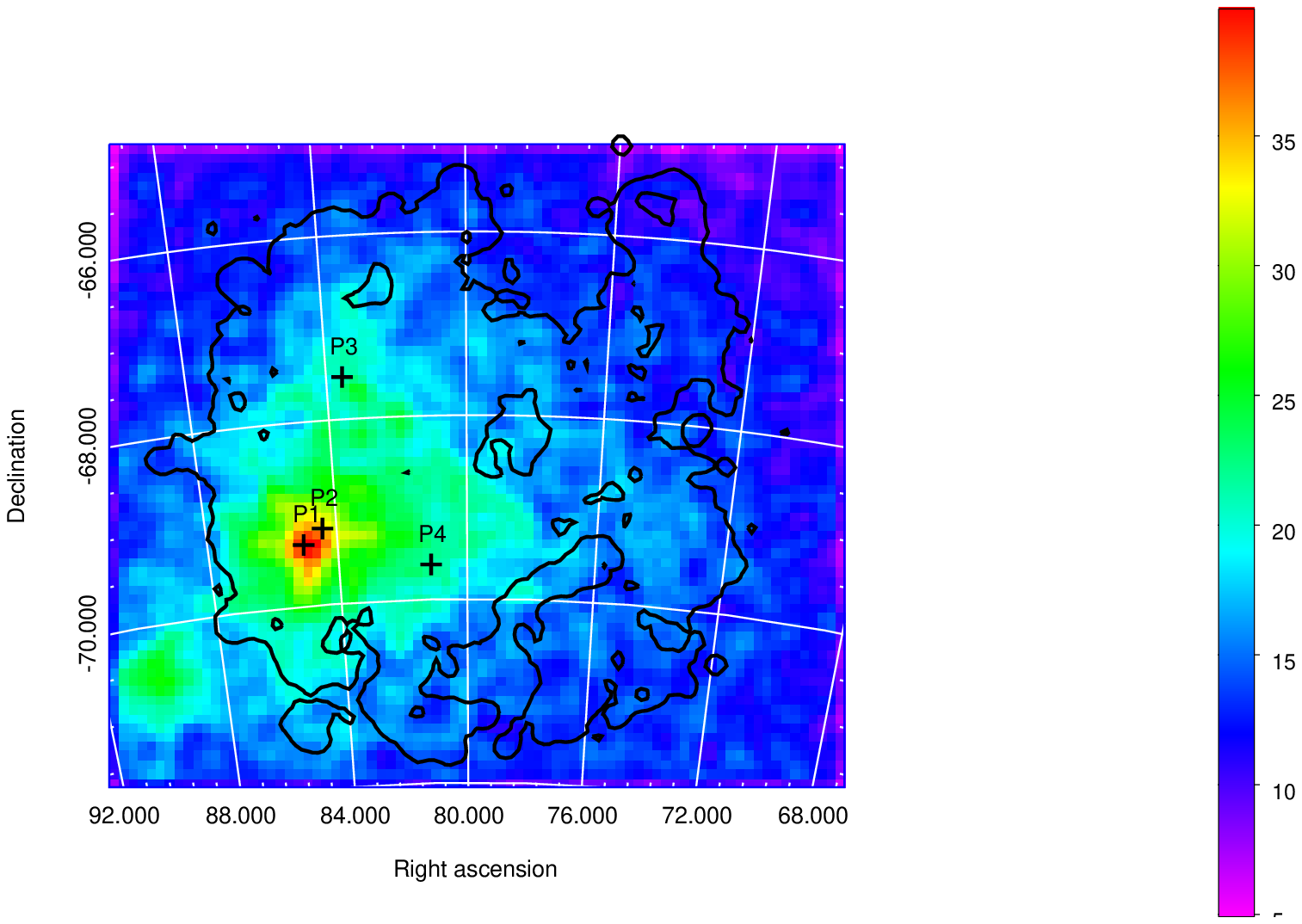}
\caption{Locations of the LMC sources in the count map of the {\it Fermi}-LAT observation of 60~MeV to 2.45~GeV, in which the crosses of P1, P2, P3 and P4 are the LMC point sources while the circles or contour are the LMC extended sources. Top: G template, the diffusion region comprises four parts, each of which is assumed to be a two dimensional Gaussian profile~\citep{2016A&A...586A..71A}; Center: D template, the LMC diffusion region makes use of the Disk and 30 Doradus, both of which employed the two dimensional Gaussian profile~\citep{2010A&A...512A...7A}; Bottom: H template, a gas model of ionized Hydrogen employing the Southern H-Alpha Sky Survey Atlas intensity distribution for the LMC diffusion region~\citep{2001PASP..113.1326G}.
\label{fig:f2}}
\end{figure}

Different templates are used for the extended sources found in the LMC field~\citep{2010A&A...512A...7A, 2016A&A...586A..71A}. Three spatial templates are considered and plotted in Fig. \ref{fig:f2}:

\begin{enumerate}

\item {\bf G template}. Two-dimensional Gaussian template model for four sources (``G1'',``G2'',``G3'',``G4''), which is called the ``analytic model'' in \citet{2016A&A...586A..71A}.

\item {\bf D template}.  A template model with the ``Disk'' and ``30 Doradus'' being modeled as a two-dimensional Gaussian. This template is used for the LMC in~\citet{2010A&A...512A...7A} and is archived in the latest {\it Fermi}-LAT extended source template catalog\footnote{\url{https://fermi.gsfc.nasa.gov/ssc/data/access/lat/4yr\_catalog/}}.

\item {\bf H template}.  A gas model of the ionized Hydrogen employing the Southern H-Alpha Sky Survey Atlas intensity distribution (H$_\alpha$) for the LMC diffusion region~\citep{2001PASP..113.1326G}. The template is also used in the
    comparative analysis of gas models~\citep{2010A&A...512A...7A,2016A&A...586A..71A}. We considered it because the gamma-ray emission of the LMC correlates better with ionized gas than that with other gases or the total gas~\citep{2010A&A...512A...7A,2016A&A...586A..71A}, which might trace the population of young and massive stars.

\end{enumerate}

\subsubsection{Photon spectral models}
\label{datamodel}
The photon models of the LMC sources depend on the selection of energy bands.
The first selection is to divide 0.06-100~GeV into 12 independent energy bands logarithmically ( i.e., performing spectral analysis on each independent energy band, hereafter the independent analysis). The second is to select a broad energy band (hereafter the broad-band analysis), which combine the several independent energy bands.

(1) For the independent analysis, we assume a single power law (PL) function to be the photon emission model of all LMC sources as used in~\citet{2013Sci...339..807A}:
\begin{equation}
F(E) = K (E/E_0 )^{-\Gamma_1}
\end{equation}
where $K$ is the normalisation, $E_0$ is the pivot energy of 200~MeV (hereafter all $E_0$ in other equations are fixed at 200~MeV) and $\Gamma_1$ is the power law index. For a narrow energy band in the independent analysis, $\Gamma_1$ is fixed at a common value of $2$~\citep{2015ApJ...808...44F}.

(2) For the broad-band analysis, several photon emission models are employed. As discussed below, the broad-band analysis is performed for the template G only. Therefore we discuss the models for the G template.
For the G1 component, we test the goodness of fit with two models, i.e., PL and Broken power law (BPL). The BPL model is given by:
\begin{equation}
F(E) = K(E/E_0) ^{-\Gamma_1} [1 +(E/E_{\rm br})^{(\Gamma_2 -\Gamma_1)/s}]^{-s}
\end{equation}
where $\Gamma_1$ and $\Gamma_2$ are the power law indices before
and after the break energy $E_{\rm br}$, and $s$ is
the smoothness of the break, which is fixed at $0.1$~\citep{2013Sci...339..807A}.

For point source P1, its photon flux can be modeled with a PL with an exponential cutoff (PLC):
\begin{equation}
F(E) = K (E/E_0 )^{-\Gamma_1}\exp(-E/E_c)
\end{equation}
where $E_c$ is the exponential cutoff energy.
Other extended sources (G2, G3, G4) and point sources (P2, P3, P4) are all modeled by a PL photon spectrum with photon index $\Gamma_1$ being free for the broad energy bands, which is different from the independent analysis of the narrow energy bands. Our model selection is consistent with that in~\citet{2016A&A...586A..71A}.

\subsection{Results}
\subsubsection{Results of the independent analysis}
In the independent analysis, we divide the LAT gamma rays between 60~MeV and 100~GeV into 12 logarithmic spaced energy bands, in each of which the spectrum is fitted by a PL photon model with a fixed photon index of $2.0$.

\begin{figure}
\centering
\includegraphics[scale=0.80]{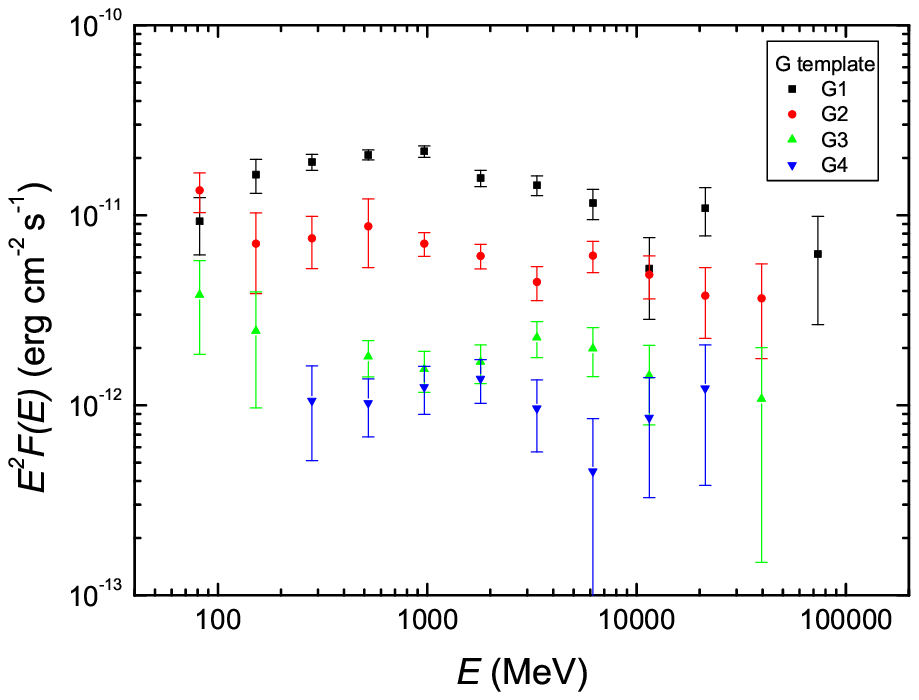}
\includegraphics[scale=0.80]{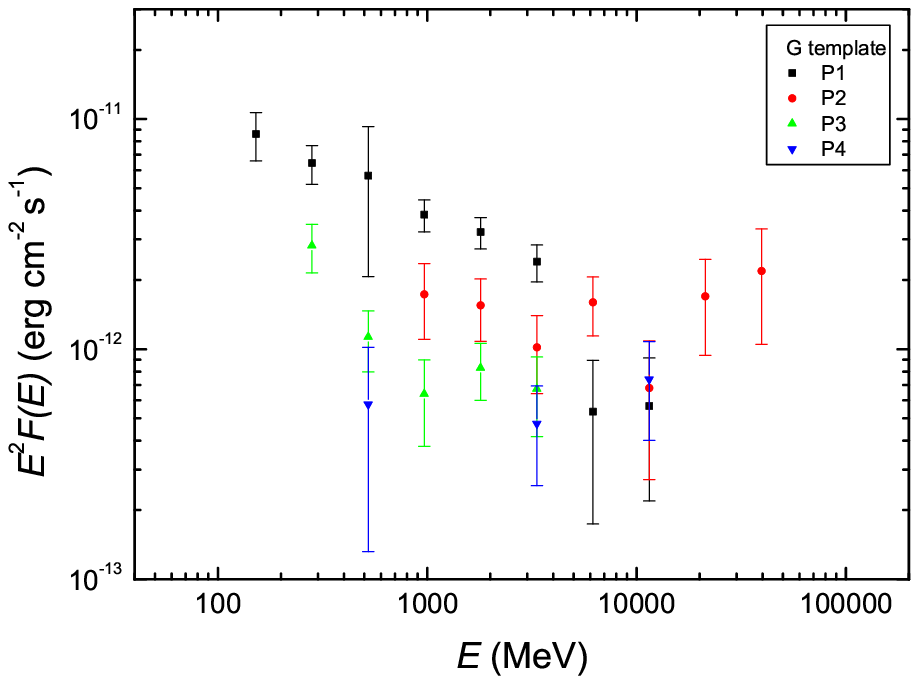}
\includegraphics[scale=0.80]{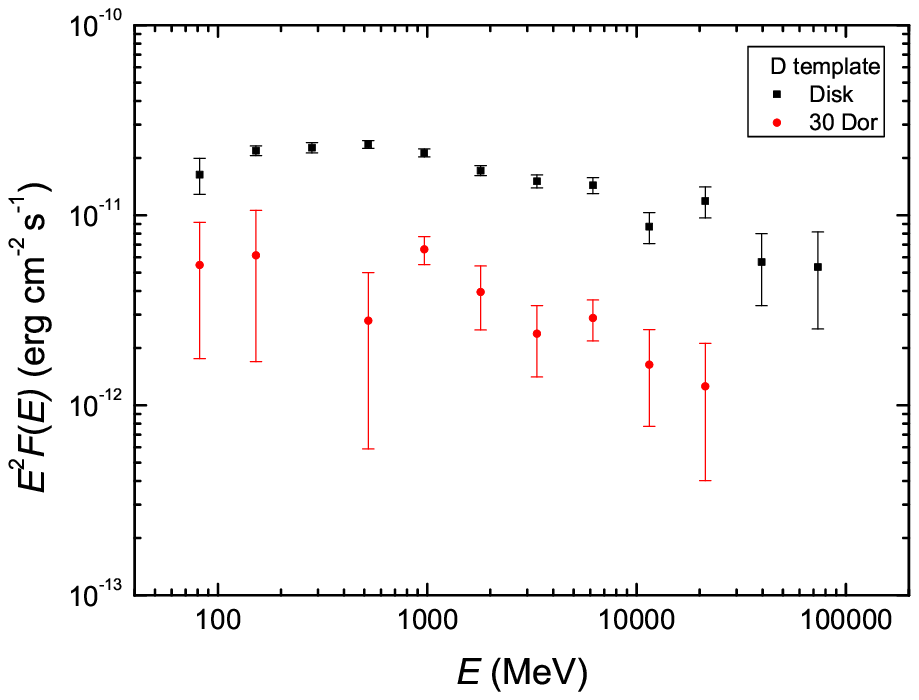}
\includegraphics[scale=0.80]{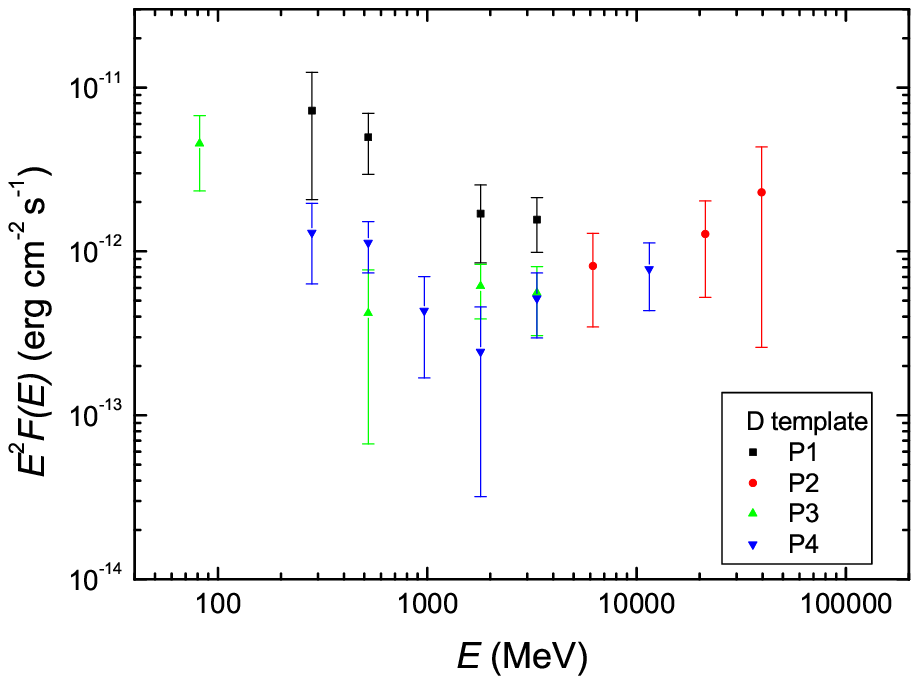}
\includegraphics[scale=0.80]{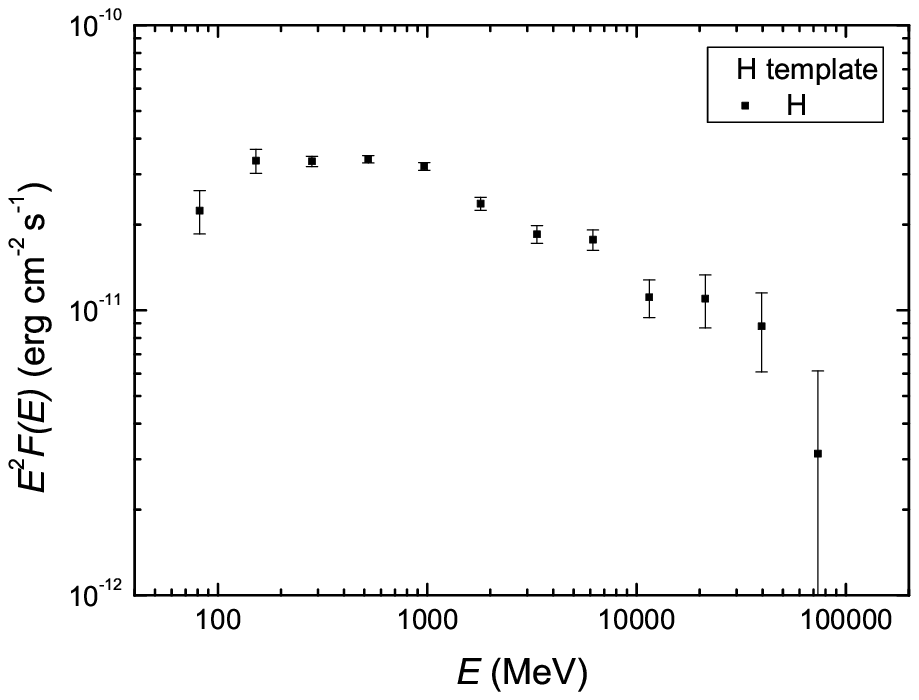}
\includegraphics[scale=0.80]{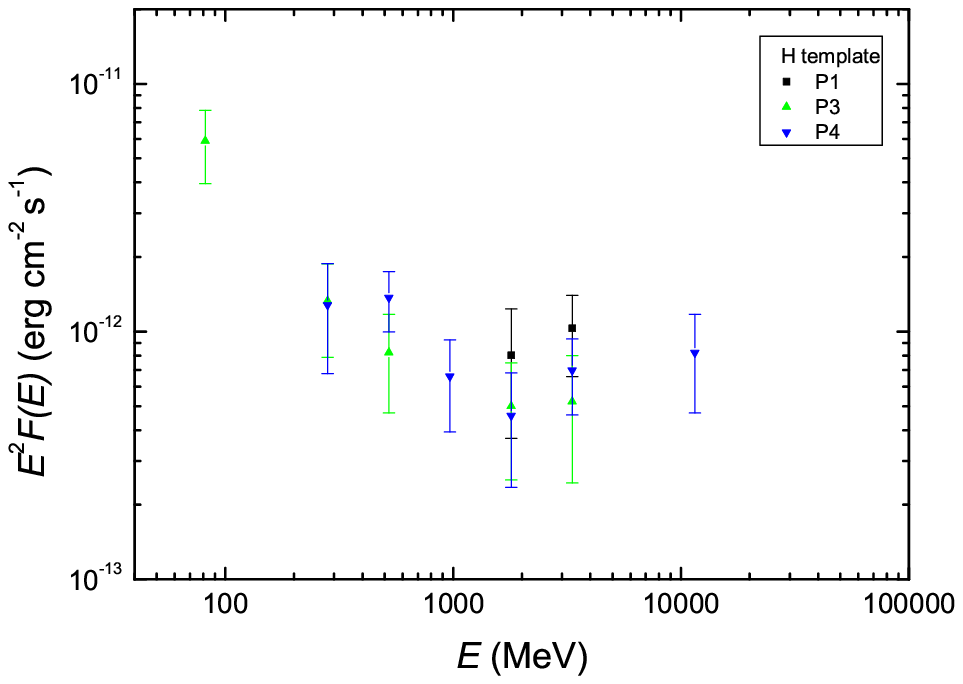}
\caption{The gamma-ray spectral data of the LMC extended components and point sources by the independent analysis. Top for the G template, middle for the D template and bottom for the H template. In order to be visible of the spectrum behavior, the upper limits are not plotted.  \label{fig:f3}}
\end{figure}

The spectral results in three templates can be found in Fig. \ref{fig:f3}, in which both the extended and point sources are plotted. As a good residual count map is obtained and the value of ``like2obj.getRetCode()'' is zero\footnote{\url{https://fermi.gsfc.nasa.gov/ssc/data/analysis/scitools/extended/extended.html}},
the fit thus is considered to be good in each independent band. The results of the large-scale disk in three templates can be found in Tab. \ref{tab:binflux}. We discuss the results for each template following:
\begin{enumerate}
\item {\bf G template}. The G template is useful to distinguish extended sources from point sources. Apparently, the 60~MeV-100~GeV spectrum of the G1 component cannot be fit by a PL function, thus we will test the fitting goodness with two functions, a PL and a BPL.  Others three extended components can be fitted by a PL function within the uncertainties. The spectrum of the point source P1 decays rapidly up to $\sim$ 4~GeV, which could be fit by a PLC function. Other three point sources can be modeled with a PL function.
\item {\bf D template}. The emission of the Disk component rises before several hundreds of ~MeV and decays up to 100~GeV. The 30 Dor component is observed in eight independent energy bands, while in other 4 energy bands no significant emission is detected from the 30 Dor. It can be explained as that the luminous point sources P1 and P2 are lying near the center of the 30 Dor. The source P1 shows an initial fast decay and then exhibit no significant emission. The source P2 is observed in three higher energy independent bands ($>$ 5~GeV). The emission of the source P3 are detected in three lower energy bands ($<$ 5~GeV). The P4 source are decomposed in 6 energy bands, which can be fitted by a PL function.
\item {\bf H template}. The emission of the H component rises quickly before 300~MeV followed by a flat spectrum behavior up to 1GeV, and then decays to 100~GeV. For the source P1, we obtained the low level emission in 2 independent energy bands. The source P3 and P4 can be significantly detected in four and six energy bands respectively, both of which show a PL decay. The emission of the source P2 is absent in this template. We found that, in the H template, the intensive region of the ionized Hydrogen is lying around the position of P1 and P2, which can account for the dim emission and non-detection of P1 and P2 respectively.
\end{enumerate}

\begin{table*}
\centering
\caption{Spectral data of the LMC large scale disk as well as the Galactic diffusion emission and the isotropic diffusion emission in the range of 60~MeV to 100~GeV measured by the {\it Fermi}-LAT. \label{tab:binflux}}
\begin{tabular}{ccccccccc}
\hline
Energy & Component &TS\tablenotemark{a} & $F_{Disk}$\tablenotemark{b} & $f_{Disk}$\tablenotemark{b} & $P_g$\tablenotemark{c} &  $N_i$\tablenotemark{d} \\
MeV &       &    & $10^{-9}{\rm ph\ cm^{-2}\  s^{-1}}$ & $10^{-11}{\rm erg\ cm^{-2}\  s^{-1}}$ &  &   \\
\hline
60	-	111	&	G1	&	47	&	72.17	$\pm$	24.12	&	0.93	$\pm$	0.31	&	1.01	 $\pm$	0.04	&	1.16	$\pm$	0.02	\\
111	-	207	&	...	&	311	&	68.80	$\pm$	13.90	&	1.64	$\pm$	0.33	&	0.93	 $\pm$	0.03	&	1.17	$\pm$	0.03	\\
207	-	383	&	...	&	476	&	43.10	$\pm$	4.23	&	1.91	$\pm$	0.19	&	0.95	 $\pm$	0.02	&	1.14	$\pm$	0.03	\\
383	-	711	&	...	&	522	&	25.30	$\pm$	1.51	&	2.08	$\pm$	0.13	&	0.90	 $\pm$	0.01	&	1.27	$\pm$	0.02	\\
711	-	1320	&	...	&	481	&	14.20	$\pm$	0.96	&	2.17	$\pm$	0.15	&	0.94	 $\pm$	0.02	&	1.18	$\pm$	0.07	\\
1320	-	2449	&	...	&	199	&	5.54	$\pm$	0.54	&	1.57	$\pm$	0.15	&	0.95	 $\pm$	0.03	&	1.19	$\pm$	0.11	\\
2449	-	4545	&	...	&	126	&	2.74	$\pm$	0.33	&	1.44	$\pm$	0.17	&	0.89	 $\pm$	0.05	&	1.31	$\pm$	0.15	\\
4545	-	8434	&	...	&	52	&	1.19	$\pm$	0.21	&	1.16	$\pm$	0.21	&	0.93	 $\pm$	0.08	&	1.34	$\pm$	0.15	\\
8434	-	15651	&	...	&	7	&	0.29	$\pm$	0.13	&	0.52	$\pm$	0.24	&	1.31	 $\pm$	0.20	&	0.95	$\pm$	0.17	\\
15651	-	29042	&	...	&	22	&	0.32	$\pm$	0.09	&	1.09	$\pm$	0.31	&	0.89	 $\pm$	0.34	&	1.06	$\pm$	0.21	\\
29042	-	53891	&	...	&	$<1$	&	$<0.08$	&	$<0.73$	&	0.49	$\pm$	0.71	&	1.34	 $\pm$	0.36	\\
53891	-	100000	&	...	&	5	&	0.05	$\pm$	0.03	&	0.63	$\pm$	0.36	&	1.45	 $\pm$	1.02	&	0.92	$\pm$	0.48	\\
			&		&		&				&				&				&				\\
60	-	111	&	Disk	&	250	&	127.00	$\pm$	27.30	&	1.64	$\pm$	0.35	&	1.01	 $\pm$	0.04	&	1.16	$\pm$	0.02	\\
111	-	207	&	...	&	914	&	91.60	$\pm$	5.41	&	2.19	$\pm$	0.13	&	0.94	 $\pm$	0.03	&	1.17	$\pm$	0.03	\\
207	-	383	&	...	&	1229	&	51.30	$\pm$	3.20	&	2.27	$\pm$	0.14	&	0.96	 $\pm$	0.02	&	1.15	$\pm$	0.03	\\
383	-	711	&	...	&	1330	&	28.60	$\pm$	1.35	&	2.36	$\pm$	0.11	&	0.91	 $\pm$	0.02	&	1.32	$\pm$	0.05	\\
711	-	1320	&	...	&	985	&	14.00	$\pm$	0.66	&	2.13	$\pm$	0.10	&	0.94	 $\pm$	0.02	&	1.28	$\pm$	0.08	\\
1320	-	2449	&	...	&	524	&	6.06	$\pm$	0.36	&	1.72	$\pm$	0.10	&	0.96	 $\pm$	0.03	&	1.28	$\pm$	0.12	\\
2449	-	4545	&	...	&	297	&	2.86	$\pm$	0.22	&	1.51	$\pm$	0.12	&	0.90	 $\pm$	0.05	&	1.43	$\pm$	0.16	\\
4545	-	8434	&	...	&	191	&	1.47	$\pm$	0.14	&	1.44	$\pm$	0.14	&	0.93	 $\pm$	0.08	&	1.41	$\pm$	0.15	\\
8434	-	15651	&	...	&	48	&	0.48	$\pm$	0.09	&	0.87	$\pm$	0.16	&	1.31	 $\pm$	0.19	&	0.98	$\pm$	0.16	\\
15651	-	29042	&	...	&	55	&	0.35	$\pm$	0.07	&	1.19	$\pm$	0.22	&	0.90	 $\pm$	0.34	&	1.10	$\pm$	0.21	\\
29042	-	53891	&	...	&	10	&	0.09	$\pm$	0.04	&	0.57	$\pm$	0.23	&	0.48	 $\pm$	0.69	&	1.36	$\pm$	0.34	\\
53891	-	100000	&	...	&	6	&	0.05	$\pm$	0.02	&	0.54	$\pm$	0.28	&	1.44	 $\pm$	1.04	&	0.97	$\pm$	0.48	\\
			&		&		&				&				&				&				\\
60	-	111	&	H	&	421	&	174.00	$\pm$	30.10	&	2.24	$\pm$	0.39	&	1.01	 $\pm$	0.04	&	1.16	$\pm$	0.02	\\
111	-	207	&	...	&	2080	&	140.00	$\pm$	13.60	&	3.35	$\pm$	0.33	&	0.94	 $\pm$	0.03	&	1.17	$\pm$	0.03	\\
207	-	383	&	...	&	2600	&	75.00	$\pm$	3.08	&	3.33	$\pm$	0.14	&	0.96	 $\pm$	0.02	&	1.14	$\pm$	0.03	\\
383	-	711	&	...	&	2928	&	41.20	$\pm$	1.23	&	3.39	$\pm$	0.10	&	0.91	 $\pm$	0.02	&	1.29	$\pm$	0.05	\\
711	-	1320	&	...	&	2587	&	21.00	$\pm$	0.65	&	3.20	$\pm$	0.10	&	0.94	 $\pm$	0.02	&	1.23	$\pm$	0.08	\\
1320	-	2449	&	...	&	1071	&	8.34	$\pm$	0.44	&	2.37	$\pm$	0.13	&	 0.96	$\pm$	0.03	&	1.27	$\pm$	0.12	\\
2449	-	4545	&	...	&	480	&	3.52	$\pm$	0.25	&	1.85	$\pm$	0.13	&	0.89	 $\pm$	0.05	&	1.45	$\pm$	0.16	\\
4545	-	8434	&	...	&	398	&	1.82	$\pm$	0.15	&	1.77	$\pm$	0.15	&	0.93	 $\pm$	0.08	&	1.42	$\pm$	0.15	\\
8434	-	15651	&	...	&	106	&	0.61	$\pm$	0.09	&	1.11	$\pm$	0.17	&	1.30	 $\pm$	0.18	&	0.98	$\pm$	0.16	\\
15651	-	29042	&	...	&	53	&	0.33	$\pm$	0.07	&	1.10	$\pm$	0.23	&	0.85	 $\pm$	0.35	&	1.17	$\pm$	0.21	\\
29042	-	53891	&	...	&	24	&	0.14	$\pm$	0.04	&	0.88	$\pm$	0.27	&	0.49	 $\pm$	0.68	&	1.33	$\pm$	0.34	\\
53891	-	100000	&	...	&	2	&	0.03	$\pm$	0.03	&	0.31	$\pm$	0.30	&	1.35	 $\pm$	1.05	&	1.08	$\pm$	0.49	\\\\

\hline
\end{tabular}
\tablenotetext{a}{The test-statistic
value (TS) is roughly equal to the squared detection significance of the corresponding component ~\citep{1996ApJ...461..396M}.}
\tablenotetext{b}{Photon flux and energy flux of the corresponding component in unit of $10^{-9}{\rm ph\ cm^{-2}\  s^{-1}}$ and $10^{-11}{\rm erg\ cm^{-2}\  s^{-1}}$ respectively.}
\tablenotetext{c}{Prefactor of the Galactic diffusion emission, which is the relative intensity to the Galactic diffuse emission template derived by {\it Fermi} team~\citep{2016ApJS..223...26A}. Typically, it is not far from 1.0.}
\tablenotetext{d}{Normalisation of the isotropic diffusion emission, which is the relative intensity to the extragalactic isotropic emission template derived by {\it Fermi} team~\footnote{\url{https://fermi.gsfc.nasa.gov/ssc/data/access/lat/BackgroundModels.html}}, which should be close to 1.0.}
\end{table*}

Among the three templates, the G template is the best one to decompose the LMC extended and point sources in the independent energy bands.
In order to obtain the spectrum of a pure large-scale disk component (G1), we perform an analysis using the G template in the sections below.

\subsubsection{Results of the broad-band analysis}
As shown in Fig. \ref{fig:f3}, the spectrum of
the independent analysis in the G template
has a rapid rise below about
500~MeV and then transits to a much softer spectrum.
To quantify the significance of the spectral break, we perform
comparative fitting in two broad-band energy ranges, i.e., 0.06-2.45~GeV and
0.06-100~GeV. The former energy range covers the 6 independent energy bands and is
close to the energy range (0.06-2~GeV) used in~\citet{2013Sci...339..807A}, in which a characteristic $\pi^0$ decay feature is reported to be found in two Galactic SNRs.
The latter energy range is selected in order to test whether the BPL is still a good function to fit the gamma-ray emission up to 100~GeV.

Given an input photon model, the probability of obtaining the data as observed is noted by $L$, which
is the product of the probabilities of obtaining the observed counts by the LAT in each bin, i.e.,
\begin{equation}
L = \prod_k \frac{m_k^{n_k}e^{-m_k}}{n_k!}=e^{-N_{pred}}\prod_k\frac{m_k^{n_k}}{n_k!}
\end{equation}
where $k$ is an index over image pixels in both space and energy,
$m_k$ indicates the number of counts predicted by the model at
pixel $k$, $n_k$ is the observed number of counts at pixel $k$, and $N_{pred}$
is the total number of observed counts
\footnote{\url{https://fermi.gsfc.nasa.gov/ssc/data/analysis/documentation/Cicerone/Cicerone_Likelihood/Likelihood_formula.html}}.

We calculate the test-statistic
value (TS) defined as $-2\log(L_0/L_1)$, where $L_0, L_1$ correspond to the likelihood value
for the case without the G1 component, with the G1 component respectively~\citep{1996ApJ...461..396M}. Since the BPL is a nested model with two additional degrees of freedom (dof) more than the PL,  a significant change can be reached when ${\rm \Delta TS}$ is larger than 25 ($\sim 5 \sigma$) from the BPL to the PL, where ${\rm \Delta TS}$ approximately follows the $\chi^2$ distribution~\citep{2013Sci...339..807A,2014MNRAS.441.3591H}.

First, we fit the spectrum between
0.06~GeV and 2.45~GeV with the PL and the BPL functions.
The BPL yields a significantly larger TS value than the PL,
with an improvement of $\Delta{\rm TS}=66$ (see Tab. \ref{tab:compare}), i.e., statistical significance of $\sim 8.1\sigma$.
The  photon index is $\Gamma_1=1.48\pm0.09$ below the break
energy of $497\pm78$~MeV, above which the
photon index is $\Gamma_2=2.35\pm0.11$.
Second, we test if the BPL model can fit the data in a border
energy range, i.e., 0.06-100~GeV. The BPL still yields a larger TS value, with an improvement of $\Delta{\rm TS} = 180$, i.e., statistical significance of $\sim 13.4\sigma$ over the PL. The  photon index is $\Gamma_1=1.39\pm0.03$ below the break
energy of $532\pm20$~MeV, above which the
photon index is $\Gamma_2=2.40\pm0.03$.
The results in both two broad energy bands show that the BPL is the better function to fit the gamma rays of the G1 component, indicating that a break at $\sim$500 MeV exists in the spectrum of the large scale disk component of the LMC.

\subsubsection{Comparative analysis without the data between 60-200~MeV}
\label{compare_above_200MeV}
To compare with the results in the former literature~\citep{2016A&A...586A..71A},
we perform the spectrum analysis on the {\it Fermi}-LAT data after removing the data of 60-200~MeV,
i.e., in the energy range of 0.2-2.45~GeV and 0.2-100~GeV. The results are shown
in Tab. \ref{tab:compare}. In the former energy range, the
PL fitting gives an photon index of about $2.0$, which is softer than that
includes the data below 200~MeV. The BPL has an improvement of
$\Delta{\rm TS} = 32$ to the PL, i.e., statistical significance of $\sim 5.7\sigma$. This, however, is lower than the improvement in the case including the data in 60-200~MeV, that is $\sim 8.1 \sigma$.

In case of 0.2-100~GeV, the BPL with a break energy of $490\pm18$~MeV is found to give a better fit than the PL. However, the change in TS of 70,  say $\sim 8.4\sigma$, is much smaller than the case including the data in 60-200~MeV, that is $\sim 13.4\sigma$.

The significant improvement of the fit when including the low energy data of 60-200~MeV  favors the existence of the $\pi^0$ decay bump in the gamma-ray spectrum of the LMC disk. The 60-200~MeV data  also provide extra flux points to constrain the physical model parameters statistically, see Sec. \ref{model_results}.

\begin{table*}
\centering
\caption{Broad-band analysis results of the G1 component. \label{tab:compare}}
\begin{tabular}{ccccccccccc}
\hline
Model  &E &Component & K\tablenotemark{a} & $\Gamma_1$\tablenotemark{b} & $\Gamma_2$\tablenotemark{c} & $E_{\rm br}$\tablenotemark{d} &$\log L_0$\tablenotemark{e} &$\log L_1$\tablenotemark{e} & ${\rm TS}$\tablenotemark{e} & $\Delta {\rm TS}$\tablenotemark{e} \\
&GeV&&&&&MeV&&&\\
\hline
PL  & 0.06-2.45&G1& $4.0 \pm 0.2$ & $1.89\pm 0.03$  & - & -   &-546725 &-546350   &  750 &-\\
BPL &...&... & $3.8 \pm 0.1$ & $1.48\pm 0.09$& $2.35 \pm 0.11$ & $497 \pm 78$ &... &-546317  &  816 & 66\\
...\\
PL  & 0.06-100&...  & $4.1 \pm 0.2$ & $2.06\pm 0.02$  & - & -   &-261700 &-261305  &  790 &-\\
BPL &...&... & $3.4 \pm 0.1$ & $1.39\pm 0.03$& $2.40 \pm 0.03$ & $532 \pm 20$ &... &-261215  &  970 & 180\\
\hline

PL  & 0.2-2.45&G1& $4.8 \pm 0.3$ & $2.01\pm 0.04$  & - & -   &-418860 &-418499   &  722 &-\\
BPL &...&... & $3.3 \pm 0.1$ & $1.21\pm 0.03$& $2.27 \pm 0.03$ & $429 \pm 10$ &... &-418483  &  754 & 32\\
...\\
PL  & 0.2-100   &...  & $5.2 \pm 0.2$ & $2.15\pm 0.02$  & - & -   &-264997 &-264589  &  816 &-\\
BPL &...&... & $3.0 \pm 0.1$ & $1.21\pm 0.04$& $2.34 \pm 0.03$ & $490 \pm 18$  &... &-264554  &  886 & 70\\
\hline
\end{tabular}
\tablenotetext{a}{\ Normalisations in unit of $10^{-10}{\rm cm^{2}\  s^{-1}\ MeV^{-1}}$.}
\tablenotetext{b}{\ Photon index of PL or BPL (pre-break).}
\tablenotetext{c}{\ Photon index of BPL (post-break).}
\tablenotetext{d}{\ Break energy of BPL.}
\tablenotetext{e}{\ The TS defined as $2\log(L_1/L_0)$, where $L_1, L_0$ correspond to the likelihood value for the the case with the G1 or without the G1. $\Delta {\rm TS}$ is the change TS from BPL to PL, which approximately follows a $\chi^2$ distribution.}
\end{table*}

\section{The physical models}
\label{model}
In this section, we explore the origin of the diffuse gamma-ray emission by the physical models. We consider two radiation models for the gamma-ray data between 60~MeV and 100~GeV, i.e., the electron bremsstrahlung model and the neutral pion decay model.

\subsection{The electron bremsstrahlung model}
In the electron bremsstrahlung model, we consider both a PL distribution $dN_e/dE_e\propto E_e^{-s_{e1}}$ and
a BPL distribution, i.e., $dN_e/dE_e= C_e(E_e/E_{b,e})^{-s_{e1}}$ and
$dN_e/dE_e= C_e(E_e/E_{b,e})^{-s_{e2}}$ below and above the break energy
$E_{b,e}$, for the injected electrons. The bremsstrahlung emission flux emitted by ultra-relativistic electrons can then be given by~\citet{1971NASSP.249.....S} and~\citet{2015ApJ...808...44F}:
\begin{equation}
E_\gamma^2 F_{brem}(E_\gamma)=E_\gamma \int_{E_\gamma}^{E_{max}}cn_H \sigma_{brem} b^{-1}(E_e)N_e(E_e)dE_e
\end{equation}
where $\sigma_{brem}=\frac{4\alpha}{\pi}\sigma_{\rm T}\ln(183)\simeq 3.22\times 10^{-26}\rm \ cm^{2}$ is the cross section, in which $\alpha$ is fine structure
constant, and $\sigma_{\rm T}$ is the Thomson scattering cross section, see Equation (23) of~\citet{2015ApJ...808...44F}.
$c$ is the speed of light and $E_{max}$ is fixed to 2~TeV in the
calculation, which results in a rollover at high energies,
improving the agreement with the GALPROP model beyond 100 GeV~\citep{2011A&A...534A..54S,2013ApJ...773..104C}.
Here $b(E_e)$ is the sum of electron energy-loss
rates by synchrotron radiation, inverse-Compton scattering,
bremsstrahlung radiation and ionization~\citep{1964ocr..book.....G, 2015ApJ...808...44F}, i.e.,
\begin{equation}
b(E_e)=b_{syn}(E_e)+b_{IC}(E_e)+b_{brem}(E_e)+b_{ion}(E_e).
\end{equation}
where $b_{syn}\varpropto B^2$ ($B$, magnetic field intensity in unit of ${\rm \mu G}$), $b_{IC}\varpropto {\rm U_{ph}}$ (${\rm U_{ph}}$, photon energy density in unit of ${\rm eV cm^{-3}}$), $b_{brem}\varpropto n_{\rm H}$ and
$b_{ion}\varpropto n_{\rm H}$ ($n_{\rm H}$, the density of hydrogen atom in unit of ${\rm \ cm^{-3}}$), see Equation (32) to (37) of~\citet{2015ApJ...808...44F}.
There are five free parameters for bremsstrahlung model with a PL injection electron distribution, i.e., $n_{\rm H}$, $B$, ${\rm U_{ph}}$, the normalisation (${\rm C}_{e}$) and the injected electron spectrum index ($s_{e1}$). As for a BPL electron distribution, two additional free parameters are considered, i.e., the post-break spectrum index ($s_{e2}$) and the break energy ($E_{b,e}$).

\subsection{The neutral pion decay model}
For the neutral pion decay model, the gamma-ray flux is calculated by the semi--analytical method proposed by~\citet{2006PhRvD..74c4018K}:
\begin{equation}
E_\gamma^2F_{\pi}(E_\gamma)=
E_\gamma^2\int_{E_\gamma}^{\infty}cn_H{\sigma}_{pp}(E_p)\frac{dN_p}{dE_p}(E_p)f_\gamma(\frac{E_\gamma}{E_p},E_p)\frac{dE_p}{E_p}
\end{equation}
where $\sigma_{pp}= 10^{-27}(34.3+1.88M+0.25M^2)(1-(1.22{\rm TeV}/E_p)^4)^2\rm \ cm^{2}$ is the cross section of proton-proton
collision, in which $M=\ln(E_p/{\rm 1TeV})$, see Equation (79) of~\citet{2006PhRvD..74c4018K}. Here $dN_p/dE_p= C_pE_p^{-s_p}$ is the
spectrum of cosmic-ray protons with $C_p$ a normalisation, and $f_\gamma$ is the spectrum of
secondary gamma rays produced in a single proton-proton collision,
with $E_p$ and $E_\gamma$ being the cosmic-ray proton energy and
the generated gamma-ray energy respectively.
There are two free parameters in this model: the proton index $s_p$ and the product (defined as $C^{\ \prime}_p$) of the normalization of the proton spectrum $C_p$ and the density of hydrogen atoms $n_{\rm H}$, since the $n_{\rm H}$ can be extracted from the integration.

\section{Modeling Result and Discussion}
\label{model_results}
\subsection{Method}
For our fitting of 11 flux points, the $\chi^2$ can be derived:
\begin{equation}
\chi^2=\sum_{i=1}^{11}{\frac{(f_{m,i}-f_{obs,i})^2}{\sigma_{f_{obs,i}}^2}}
\end{equation}
where $f_{m,i}$ is the predicted flux by the physical model, $f_{obs,i}$ is the LAT-observed flux ($E^2F(E)$) in the the $i$th energy bin with corresponding error of $\sigma_{f_{obs,i}}$. A $\chi^2$ comparable with the degrees of freedom (dof) is considered as an acceptable fit, i.e., the reduced $\chi^2$ (labeled as $\chi_r^2$) is between 0.75 and 1.50~\citep{2011ApJ...730..141Z}. After deriving a best fit, the resultant $\chi^2$ is labeled as $\chi^2_{\rm best}$. The $1 \sigma$  error of a parameter is calculated by $\chi^2_{\rm error}=\chi^2_{\rm best}+\chi^2_{1 \sigma}$ while other parameters are fixed at the best-fit values. $\chi^2_{1 \sigma}$ can be calculated by integrating the $\chi^2$ probability density function of the corresponding degrees of freedom to 1$\sigma$.

In our following analysis, we also consider that if the resultant parameter values are consistent with values from other papers or observations (hereafter reference values). For example, when Bremsstrahlung losses dominates the gamma-ray emission, $n_{\rm H}$ can be high as
$2{\rm \ cm^{-3}}$~\citep{2003ApJS..148..473K}. The inverse Compton losses are not important and thus we consider a low photon energy density of ${\rm U_{ph}=0.57\ eV cm^{-3}}$~\citep{2010A&A...519A..67I}, where we also set a lower boundary of ${\rm U_{ph, limit}=0.01\ eV cm^{-3}}$. For the magnetic field strength, $B$ could be in the range of ${\rm 2 - 7\ \mu G}$~\citep{2005Sci...307.1610G,2010A&A...512A...7A,2012ApJ...759...25M,2015ApJ...808...44F}.

\subsection{Modeling Results}

\begin{figure}
\centering
\includegraphics[scale=0.50]{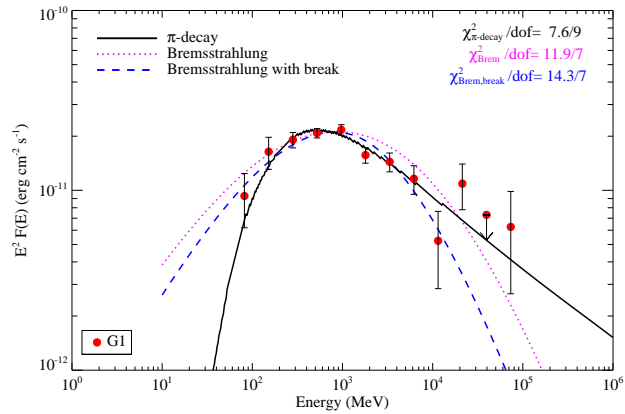}
\caption{Physical modeling of the G1 component. The red points are the energy fluxes of the G1 component measured by the {\it Fermi}-LAT, in which the errors are statistical only and the upper limit is at 95\% confidence level. The solid line is the modeling result of the neutral pion decay. The dotted line is for the Bremsstrahlung model with a PL electron spectrum, whose power index is fixed at $2.0$. The dashed line is for the Bremsstrahlung model with a BPL electron spectrum, whose distribution is same as that of our Galaxy but with a parameterized normalisation~\citep{2015ApJ...808...44F}. The derived parameter values of these models are present in Tab. \ref{tab:physical}.   \label{fig:f4}}
\end{figure}

Bremsstrahlung with the PL injected electron spectrum. First, all parameters are unfixed. The results can be found in Tab. \ref{tab:physical}. This fit is acceptable with $\chi_r^2=1.28$. However, the resultant electron spectrum index of $1.39_{+0.12}^{-0.11}$ is much harder than that in our Galaxy, i.e., $2.0-2.4$~\citep{1997JPhG...23.1765P}. Then we allow the electron spectrum index to vary between $2.0-2.4$ and find that smaller values of the electron spectrum index will result in  smaller $\chi_r^2$ close to $1$, which means a good fit. Fixing the electron spectrum index of $2.0$,  it leads to a bit worse goodness of $\chi_r^2=1.70$, but gives the constrains on all parameters, i.e., $n_{\rm H}=1.14_{+0.10}^{-0.10}\ {\rm \ cm^{-3}}$, $B=4.94_{+0.34}^{-0.31}\ {\rm \mu G}$, ${\rm U_{ph}}=0.81_{+1.18}^{-0.62}\ {\rm eV cm^{-3}}$, which are comparable to the reference values. The fit with fixed electron spectral index is considered and plotted in Fig.~\ref{fig:f4}.

Bremsstrahlung with the BPL injected electron spectrum. Initially all parameters are allowed to float. The photon energy density is attacking the lower boundary in this fitting. Thus we fixed it at 0.01 and fit the data again. The fit is good to constrain all parameters and $\chi_r^2=1.33$. We note that the magnetic field strength ($B$) of $\sim 0.08 {\rm \ \mu G}$ is much lower than the reference value, i.e, $2-7 {\rm \mu G}$ as discussed above.
In addition, we use the same electron spectrum as that in our Galaxy, which is also
used for the LMC in~\citet{2015ApJ...808...44F}, i.e., $s_{e1}=1.80, s_{e1}=2.25$ and $E_{b,e}=4$~GeV. This fit is a bit worse, i.e., $\chi_r^2=2.04$. The derived parameters, i.e., $n_{\rm H}=1.43_{+0.14}^{-0.13}\ {\rm \ cm^{-3}}$, $B=4.84_{+0.35}^{-0.30}\ {\rm \mu G}$, ${\rm U_{ph}}=0.60_{+1.34}^{-0.69}\ {\rm eV cm^{-3}}$, are comparable to the reference values. Thus the fit with same electron spectrum distribution as that of our Galaxy is adopted and plotted in Fig.~\ref{fig:f4}.

In the pion decay model,  the resultant value of chi-square, i.e., $\chi_r^2=0.85$, implies a reasonable fit to the data. The best-fit value of the proton index ($s_p$) is $2.45_{+0.14}^{-0.13}$, which is consistent with the proton index (2.4) obtained by~\citet{2015ApJ...808...44F}.  Fig. \ref{fig:f4} shows the result of the pion decay model. Without fixing other parameters, the pion decay model thus is a preferred model to model the gamma-ray emission from the G1 component with an accepted $\chi_r^2$ value.

\begin{table*}
\centering
\caption{Derived parameters from the physical models for the G1 component. \label{tab:physical}}
\begin{tabular}{ccccccccrc}
\hline
Model & ${n_{\rm H}}$ &$B$ &${\rm U_{ph}}$ & $s_{e1}$\tablenotemark{a} & $s_{e2}$\tablenotemark{a} & $E_{e,b}$\tablenotemark{a}  & $s_p$\tablenotemark{b} & $\chi^2/$dof &$\chi_r^2$\ \tablenotemark{c}\\
&  ${\rm cm^{-3}}$ &${\mu G}$  &${\rm eV cm^{-3}}$    &    &   & MeV  &   &  & \\
\hline
Bremsstrahlung &$0.39_{+0.03}^{-0.03}$ &$2.99_{+0.20}^{-0.17}$ &$7.80_{+6.73}^{-3.15}$     & $1.39_{+0.12}^{-0.11}$   & -     & -         & -       &  7.7/6  &1.28\\
... &$1.14_{+0.10}^{-0.10}$ &$4.94_{+0.34}^{-0.31}$ &$0.81_{+1.18}^{-0.62}$     & 2.00(fixed) & -  & -         & -             &  11.9/7  &1.70\\
\hline
Bremsstrahlung with Break  &$2.59_{+0.19}^{-0.19}$ &$0.08_{+0.004}^{-0.003}$   &0.01(fixed)   & $1.45_{+0.42}^{-0.74}$ & $2.41_{+0.06}^{-0.06}$  & $1318_{+442}^{-382}$ & -   &  6.7/5   &1.33\\
...  &$1.43_{+0.14}^{-0.13}$ &$4.84_{+0.35}^{-0.30}$   &$0.60_{+1.34}^{-0.69}$   & 1.80(fixed) & 2.25(fixed)  & 4000(fixed) & -   &  14.3/7   &2.04\\
\hline
$\pi^0$ decay   &- &-     & - & -              & -             & -               & $2.45_{+0.14}^{-0.13}$  &  7.6/9  &0.85\\
\hline
\end{tabular}
\tablenotetext{}{{\bf Note} (1) Top panel: Bremsstrahlung with injection PL electron spectrum, fit in top with all parameters free while in bottom fixing the electron spectrum index of 2.0; (2) Middle panel:  Bremsstrahlung with injection BPL electron spectrum, fit in top with all parameters free (the ${\rm U_{ph}}$ attacks the lower boundary, so we fix it at 0.01, see text) while in bottom fixing the electron spectrum parameters similar to our Galaxy but with a parameterized normalisation~\citep{2015ApJ...808...44F}; (3) Bottom panel: $\pi^0$ decay model with all parameters free. All errors are at 1 $\sigma$.}
\tablenotetext{a}{\ Electron energy spectral index and/or break energy of the injection electron spectrum.}
\tablenotetext{b}{\ $s_p$ is the proton energy spectral index for Pion decay model.}
\tablenotetext{c}{\ The reduced $\chi^2$, generally a fit is acceptable when $\chi_r^2$ between 0.75-1.50~\citep{2011ApJ...730..141Z}.}
\end{table*}

\section{Discussion and Conclusion}
\citet{2010ApJ...709L.152A} first notice that the gamma-ray emission of the LMC, as detected by the
{\it Fermi}-LAT, is likely diffuse, i.e., it consists of two diffusion regions, Disk and 30 Doradus. \citet{2015ApJ...808...44F} re-analyze the data by
employing several combinations of the ionizing gas (H$_\alpha$) and 160$\mu m$ radiation. They find that the leptonic processes also contribute to
the gamma-ray emission of the LMC, i.e., about 3\% of the Disk (excluding 30 Doradus) gamma-ray flux is from inverse Compton and 18\% is from Bremsstrahlung. Employing the high energy photons above 200~MeV with 6 flux points, they find a proton spectrum index of $2.4$.

After subtracting the bright LMC point sources detected by {\it Fermi}-LAT~\citep{2015Sci...350..801F}, four diffusion components
are decomposed from the LMC region in an emissivity template~\citep{2016A&A...586A..71A}.
In this template, they suggest the different origins for these four decomposed diffusion components, i.e., E0, E2, E4 and E1+E3. The emissions from the large-scale disk (E0 component, largely overlapping with the G1), is likely dominated by hadronic process while others are likely of leptonic origins. For example, the E2 (largely overlapping with the G3) and the E4 (largely overlapping with the G4) could originate from the inverse-Compton process. The E1+E3 component (largely overlapping with the G2) is more favorable for the leptonic origin.
Without considering the data below 200~MeV, the spectrum of the E0 is not visible of the $\pi^0$ decay feature.

In this work, we analyzed the high-energy gamma-ray spectra of the large-scale disk in the LMC, including the data between 60-200~MeV that was not considered in previous works.
We decomposed a large-scale disk, i.e., the G1 component, from other spatial components in the LMC, and, for the first time, found a spectrum break around 500~MeV for the disk.
The obtained gamma-ray emission can be well reproduced by the pionic gamma rays from $pp-$collision between the gas in the LMC disk and protons with a bit harder spectrum than that in our Galaxy, while the bremsstrahlung emission is marginally consistent with the observed spectrum. We conclude that, the current {\it Fermi}-LAT data of the LMC large scale disk emission favors a hadronic origin, although a leptonic model cannot be ruled out completely.

\section*{Acknowledgments}

We thank the anonymous referee and editor for helpful comments.
We are grateful to John Gaustad and A. Hughes for discussion
with the radio map of the LMC and Francesco Capozzi for revision.
TQW thanks the hospitality of The Center for Cosmology
and AstroParticle Physics (CCAPP) in The Ohio State University.
This work is supported by the 973 program
under grant 2014CB845800, the Natural Science Foundation of China
under grants 11547029, 11625312 and
11033002, the Youth Foundation of Jiangxi Province (20161BAB211007)
and the China Scholarship Council.
PHT is supported by NSFC grants 11633007 and 11661161010.

\facility{\textit{Fermi}}.
\software{Fermi Science Tools package (v10r0p5) (https://fermi.gsfc.nasa.gov/ssc/data/analysis/documentation)}

\end{document}